\documentclass{aa}  

\usepackage{graphicx}
\usepackage[varg]{txfonts}
\usepackage{siunitx}  
\usepackage[english]{babel} 
\DeclareSIUnit\solarmass{M\ensuremath{_\odot}} 
\DeclareSIUnit\solarradius{R\ensuremath{_\odot}} 
\DeclareSIUnit\solarlum{L\ensuremath{_\odot}} 

\newcommand{\iron}{Fe K$\alpha$\ }
\usepackage{amsmath}  
\usepackage{natbib}  
\bibpunct{(}{)}{;}{a}{}{,}
\usepackage{hyperref} 
\usepackage{chapterbib} 
\usepackage{aas_macros}  
\usepackage{makeidx}       
\usepackage{comment}
\usepackage{footnote}  
\usepackage{amssymb} 
\usepackage{subfig}
\usepackage{textcomp} 
\usepackage{lmodern} 
\usepackage{caption} 
\usepackage{bm} 
\begin{document}

   \title{XMM-Newton spectrum of the radio-loud quasar 3C 215: slim accretion disk or SMBH binary?}

   \author{A. Mei
          \inst{1,2,3}
          \and
            F. Tombesi
            \inst{2,4,5,6,7}
          }

   \institute{ Gran Sasso Science Institute (GSSI), Viale Francesco Crispi 7, I-67100, L’Aquila (AQ), Italy\\
              \email{alessio.mei@gssi.it}
           \and
          Department of Physics, University of Rome “Tor Vergata”, Via della Ricerca Scientifica 1, 00133 Rome, Italy\\
          \and
	         INFN-Laboratori Nazionali del Gran Sasso, I-67100, L'Aquila (AQ), Italy\\
        \and
            Department of Astronomy, University of Maryland, College Park, MD 20742, USA\\
        \and
             NASA/Goddard Space Flight Center, Code 662, Greenbelt, MD 20771, USA\\
        \and
            INAF-Osservatorio Astronomico di Roma, Via Frascati 33, 00078 Monteporzio Catone, Italy\\
         \and
         	INFN-Tor Vergata, Via della Ricerca Scientifica 1, 00133 Rome, Italy\\
         		}

   \date{}

  \abstract
   {Radio-loud active galactic nuclei (RL AGNs) show very powerful jet emission in the radio band, while the radio-quiet (RQ) ones do not. This RL-RQ dichotomy would imply a sharp difference between these two classes, but modern theoretical models and observations suggest a common nuclear environment possibly with different working regimes.}
   {We want to explore the geometrical structure and mutual interactions of the innermost components of the broad line radio galaxy (BLRG) 3C 215, with particular interest in the accretion and ejection mechanisms involving the central supermassive black hole (SMBH). We compare these observational features with the ones of the RQ Seyfert 1 galaxies. Investigating their differences it is possible to understand more about the jet launching mechanisms, and why this phenomenon is efficient only in a small fraction of all the AGNs. }
   {Using high quality data from a $\sim60\ \si{ks}$ observation with \textit{XMM-Newton}, we carried out a detailed X-ray spectral analysis of 3C 215 in the broad energy range $0.5-10\ \si{keV}$. We modeled the spectrum with an absorbed double power-law model for the primary continuum, reprocessed by reflection from ionized and cold neutral material and modified by relativistic blurring. We also compared our results with the ones obtained with previous multi-wavelength observations.}
   {We obtain a primary continuum photon index from the corona $\Gamma_1=1.97\pm0.06$ and evidence of a jet contribution, modeled as a power law with photon index $\Gamma_2\simeq1.29$. The reflector, possibly the accretion disk and portions of the broad-line region (BLR), is ionized ($\log\xi=2.31_{-0.27}^{+0.37}\ \si{erg\ s^{-1}\ cm}$) and relatively distant from the SMBH ($R_{in}>38\ R_g$), where $R_g=GM_{BH}/c^2$ is the gravitational radius. The obscuring torus seems patchy, dust-poor and inefficient, while the jet emission shows a twisted and knotted geometry. We propose three scenarios in order to describe these characteristics: 1.) ADAF state in the inner disk; 2.) Slim accretion disk; 3.) sub-pc SMBH binary system (SMBHB). }
   {While the first scenario is not in agreement with the SMBH accretion regime, the slim disk scenario is consistent with the observational features of this radio galaxy, showing that 3C 215 is similar to non-jetted AGNs, accreting at a high rate. Nonetheless, the first two scenarios are unable to account for the particular shape of 3C 215 jet emission. The SMBHB scenario seems to be in agreement with almost all 3C 215 observational features, but we are not able to unequivocally select this source as a strong SMBHB candidate, requiring further analyses. }

   \keywords{galaxies: active –-
             galaxies: nuclei –-
             galaxies: supermassive black holes –-
             accretion, accretion disks 
               }
\titlerunning{XMM-Newton spectrum of the radio-loud quasar 3C 215}
\maketitle
\section{Introduction}
It is now widely accepted that at the center of virtually all galaxies resides a supermassive black hole (SMBH)  with mass in the range between $10^6$ and $10^9\ \si{\solarmass}$. Through the accretion of matter it is able to efficiently convert gravitational energy, powering the plasma in its sphere of influence and in relatively few cases ($\sim 1 \%$) forming an active galactic nucleus (AGN). These sources are characterized by a high luminosity and broad-band emission, ranging from the radio band to the X-rays and $\gamma$-rays, and significant time variability. This places AGNs among the brightest and most complex sources in the Universe \citep{Padovani2017}.\\
According to the Unified Scheme, their peculiar emission is produced by the interaction between the central accreting SMBH and some fundamental components like the accretion disk, the hot corona, the broad line region (BLR), the obscuring torus and outflows, which are considered ubiquitous in AGNs \citep{urry&padovani1995}. A small fraction of all the AGNs ($\sim10\%$) shows an intense emission in the radio band compared to the optical. Observations reveal that these radio-loud (RL) AGNs host a powerful jet emission originating in the nuclear regions, while for the radio-quiet (RQ) ones this is several orders of magnitude weaker or absent \citep{padovani2017b}.\\
In order to produce high energy photons and relativistic particles in jets, an interaction between the magnetic field and the accretion disk is thought to be taking place very near to a highly spinning SMBH \citep{blandford&znajek1977,ballantyne2007}, and apparently these conditions are fulfilled only in the RL sources. The exact jet-launching mechanisms are still unclear and matter of debate, making RL AGNs very interesting sources to observe and analyze. Nonetheless, this is a challenging task, being these sources intrinsically rare. For this purpose the X-ray analysis of broad line radio galaxies (BLRGs) constitute an important benchmark for the study of emission and accretion mechanisms taking place in their innermost regions, near the SMBH. This class of RL AGNs is relatively luminous in the X-rays, and they are observed with an intermediate inclination angle, large enough to not observe spectra totally dominated by the jet emission (as in the blazar class) and small enough to avoid obscuration from the torus.\\
By analyzing similarities and differences between BLRGs and their RQ counterparts, the Seyfert 1 galaxies, which show a very similar spectrum but absence or faint jet component, it is possible to explore which could be the jet production mechanisms and test the so called RL-RQ dichotomy. 
RL AGNs seem to have a smaller accretion rate \citep{panessa2007} and higher BH masses \citep{laor2000, chiaberge2011} with respect to the RQ AGNs. Moreover, RL AGNs are mostly hosted in elliptical or bulge-dominated galaxies, while RQ can be found both  in spirals and elliptical galaxies \citep{floyd2004}. This suggest a connection between RL AGNs and major merger events of the host galaxies during their evolution \citep{ravindranath2002,chiaberge2015}.\\
As pointed out by \cite{eracleous2000} BLRGs show weaker X-ray reflection signatures  with respect to Seyfert 1 galaxies, with narrow or not detectable iron lines and weak Compton reflection hump, possibly due to a particular geometry of the corona-disk system.
In fact if in the innermost region of the accretion disk there is absence of matter (i.e a truncated disk) or an advection dominated accretion flow, i.e. an ADAF disk \citep{narayan&yi1994,narayan&yi1995,narayan&mcclintock2008,abramowicz&fragile2013}, the solid angle over which the outer standard thin disk is illuminated is relatively small, leading to an overall weakening of the reflection features. \\
\cite{ballantyne2002} state that also the reprocessing from a highly ionized accretion disk ($\xi \sim 4000\ \si{erg\ cm\ s^{-1}}$) can suppress the reflected spectrum, arguing that this phenomenon is not strictly correlated to a particular geometry and that also a standard thin accretion disk can account for this.\\
There are also evidences that BLRGs have harder X-ray spectra with respect to their RQ counterparts \citep{grandi2006}. Nevertheless \cite{sambruna1999} have shown only a weak indication of a flatter spectrum in a BLRG.\\
Two important parameters that can have a pivotal importance in testing the radio-loud/radio-quiet dichotomy are the Eddington parameter $\lambda=L_{bol}/L_{Edd}$ (where $L_{bol}$ is the bolometric luminosity and $L_{Edd}$ is the Eddington luminosity) \citep{ballantyne2007} and the dimensionless black hole spin parameter $a=cJ/GM^2$. Recent observations indicate a direct proportional relation between the jet power and the spin parameter, suggesting that rapidly spinning SMBHs are more likely to form powerful jets \citep{moderski1998}.\\
\cite{garofalo2010,garofalo2014} propose a different theoretical framework, arguing that the jet formation mechanism is not triggered by the intensity of the SMBH rotation, but the mutual rotational directions between the black hole and the accretion flow, favoring the formation of jets in case of retrograde rotation ($a<0$). This may explain the observational evidence of a low fraction of RL AGN at low redshift. In addition merger episodes, which cause an high accretion rate and retrograde flows, can explain how RL AGNs are formed and why they are commonly observed in elliptical galaxies, that have a high probability of having experienced at least one significant merger event in their past evolution.\\
The jet formation mechanism can be described by a disk+jet model, linked with a jet duty-cycle. According to this model the accretion disk, which is initially intact down to the innermost stable circular orbit (ISCO), is internally disrupted because of disk instabilities and ejects matter that constitutes the jet and outflows. After this, the accretion disk refills and the cycle starts again.\\
This model, proposed by \cite{lohfink13}, is in agreement with their multi-epoch observations of a BLRG caught both in the ejection and refilling stages, and make it possible to consider BLRGs observed in different stages of their jet cycle, hence showing different observational features.\\
Recent models, supported by new X-ray observations, claim that the nuclear regions of BLRGs and Seyfert 1 are composed by very similar components that interact with the same physical phenomena but at different regimes \citep{ballantyne2014,ursini2018}.\\

\section{3C 215}\label{sec:3c215}
\begin{figure}
	\centering
	\includegraphics[width=0.5\linewidth]{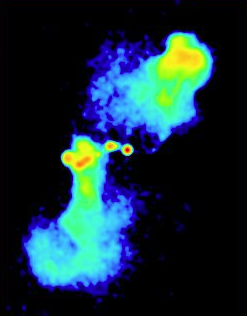}
	\caption{VLA A+B image of 3C 215 at $4860\ \si{MHz}$ \citep{bridle1994}.}
	\label{fig:3c215vla}
\end{figure}

\begin{figure}
	\centering
	\includegraphics[width=0.7\linewidth]{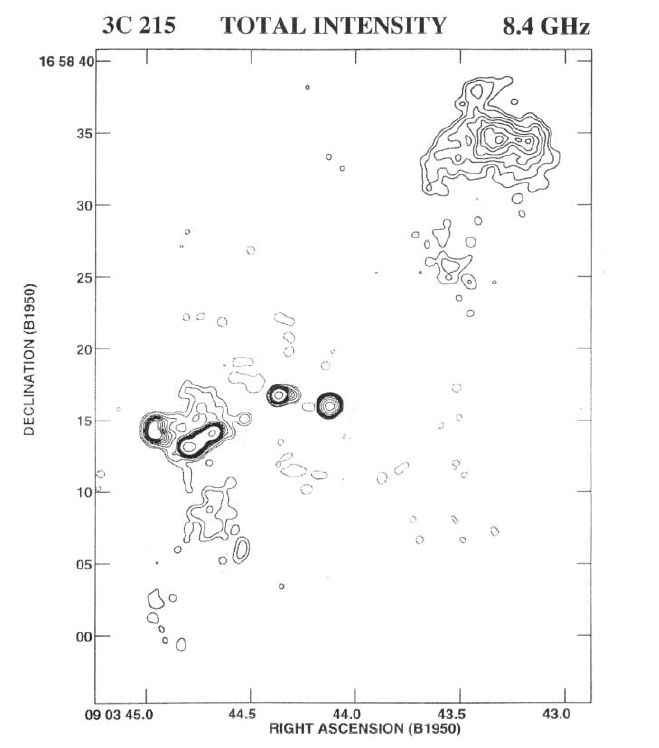}
	\caption{VLA B contour map of 3C 215 at $8.4\ \si{GHz}$ \citep{fernini2007}.}
	\label{fig:3c215vla8ghz}
\end{figure}

\object{3C 215} is a BLRG with a cosmological redshift $z=0.412$, measured through optical emission lines. This source has been observed different times in the past years, covering a large fraction of the electromagnetic spectrum, but with little information in the X-ray band. In the following we will describe the most important characteristics of 3C 215 obtained in the literature.\\
The 3C 215 host galaxy is elliptical \citep{lehnert1999} and it is probably interacting with a companion galaxy at $\sim 28\ \si{kpc}$, in a clustered environment of radius $\sim30\ \si{arcsec}$ containing at least 14 other galaxies \citep{marquez1999}. Therefore, it is likely that during its lifetime this galaxy experienced at least one major merger episode, which led the gas in the disk to be expelled and/or accreted.\\ \cite{tang2012} observed the optical/UV emission lines and measured their flux, equivalent width (EW) and full width at half maximum (FWHM) for broad and narrow emission lines in this wave-band, notably the C$_{\si{IV}}$, Mg$_{\si{II}}$ and H$_{\si{\beta}}$ emission lines. They computed three different estimates of the black hole mass $M_{BH}$ through empirical mass scale relations \citep{vestergaard2006,vestergaard2009}, adopting the median value for the black hole mass $\log M_{BH}=8.92\pm0.5$.\\
\cite{labita2006} observed a sample of Quasi-stellar Objects (QSOs), including 3C 215, in the UV band with the Hubble Space Telescope (HST) measuring the C$_{\si{IV}}$ line width and the host galaxy luminosity. They computed the black hole mass using the $M_{BH}-L_{bulge}$ relation \citep{bettoni2003} and, comparing this mass estimate with the virial theorem, they calculated the geometry factor of the BLR $f= 1/3$, indicating a disk-like BLR geometry.\\
\cite{decarli2008} further analyzed the previous sample, considering also the H$_\beta$ line. Unfortunately 3C 215 was excluded from the sample because of its low S/N, but what they found for the sample is a stratified BLR, with an inflated disk-like region that produces the C$_{\si{IV}}$ emission line and an isotropic region that produces the H$_\beta$ emission line, similar to what was found by \cite{labita2006} with the C$_{\si{IV}}$ line.\\
\cite{decarli2010} analyzed the same sample with different observatories, and obtained an independent black hole mass trough the virial theorem. They obtained the same $M_{BH}$ estimate of \cite{tang2012}, so this is the value that we will consider in this work.\\
\cite{runnoe2013} carried out a SED analysis from the IR to the X-ray band, obtaining important parameter values related to this source like the bolometric luminosity $\log L_{bol}\simeq 45.77\ \si{erg/s}$, the spectral index between the optical and the X-ray band $\alpha_{ox}\simeq -1.04$, the observable ratio between mid IR and bolometric luminosity $\log R\simeq -1.521 $ and the obscuring torus covering fraction $c\simeq 64\%$ estimated using the IR flux.\\
\cite{reeves2000} (R\&T00) and \cite{hardcastle2006} (H06) both carried out with different telescopes an X-ray analysis of 3C 215 in the $2-10\ \si{keV}$ band, but with the low S/N data available to them they could just measure the luminosity in the X-ray band $L_X$ and the photon index $\Gamma$. Their work is discussed in more detail in Sec. \ref{sec:xrayvariab}.\\
The radio band observations of 3C 215 are ambiguous \citep{bridle1994,gilbert2004,fernini2007}. This source lies in an highly dense clustered environment, and shows a knotty and twisted jet, tilted by almost $90$\textdegree\ with respect to the lobes (Fig. \ref{fig:3c215vla} and Fig.\ref{fig:3c215vla8ghz}).\\
The plume structure in the southern lobe is more similar to a Fanaroff-Riley (FR) I source while the northern lobe resembles FR II features. The particular environment of 3C 215 is probably one of the reasons of this particular behavior, and it can interfere with the jet at scales of $\sim 10\ \si{kpc}$. However, to account for this complex structure, there has to be several additional phenomena happening in the inner region of the AGN where the jet is formed, like a warped accretion disk or a secondary SMBH.\\
\section{Data reduction}\label{sec:datareduction}
3C 215 was observed with \textit{XMM-Newton} in April 17-18 2012. In this work, in order to acquire a better S/N, we used data from the EPIC-pn, EPIC-MOS1 and EPIC-MOS2 detectors with  total exposures of $59.6\ \si{ks}$ and $61.2\ \si{ks}$ for the pn and MOS cameras, respectively. Data reduction and analysis were carried out using the SAS standard procedure. We checked for the presence of soft proton flaring events in the dataset with energies between $E=10-12\ \si{keV}$. We produced a light curve in the same energy range, excluding 
all the time intervals with a count rate that exceeds the values $0.7, 0.3$ and $0.25$ for the EPIC-pn, MOS1 and MOS2 dataset, respectively.\\ Afterwards, we defined a circular region that encircles the source, with radii of $40"$ and $44"$ in the EPIC-pn and MOS CCDs, respectively. We repeated the same procedure with the background region, considering different portions of the same CCDs with a uniformly distributed number of photons for each pixel. In addition, we tested the presence of pile-up effects, excluding data contamination.\\
Using the SAS tools we produced redistribution matrix files (RMFs), ancillary response files (ARFs) and cleaned X-ray spectra in the energy range $E=0.5-10\ \si{keV}$, grouped with a minimal threshold of 25 photons per  bin. 
\section{Spectral analysis}\label{sec:spectralanalysis}
To analyze the spectrum obtained through the data reduction we used the package XSPEC (v. $12.10.1$) from the software Heasarc (v. $6.26.1$), that allowed us to analyze the spectrum using a standard spectral fitting method with $\chi^2$ statistics. The errors reported in this analysis are computed with a $1\sigma$ confidence level, while the upper/lower limits are reported with a $90\%$ confidence level.\\
Since we are using a combined spectrum produced from data of three different cameras, we took into account the cross-calibration differences between the EPIC-pn and EPIC-MOS devices through a cross-calibration parameter $ k $. Fixing $k=1$ for the EPIC-pn spectrum, we found from the EPIC-MOS spectral fit $k_{mos}=1.011$, that from now on will be considered fixed. This indicates a difference of only $\sim1 \%$ in the absolute flux between the EPIC-pn and MOS cameras, and it is consistent with the latest ESA calibration.
\subsection{Absorbed power-law}
Because of the Compton scattering interaction between the accretion disk photons and the hot corona, the X-ray spectrum of this source can be modeled as a power-law with slope given by the photon index $\Gamma$, which is absorbed by the interstellar medium (ISM) in the Milky Way with a cross-section given by the Tuebingen-Boulder model \citep{wilms2000}, called {\fontfamily{lmtt}\selectfont TBabs} in XSPEC. To compute the ISM column density we used the nH tool provided by Heasarc, which estimates for 3C 215 $N_H=3.5 \cdot 10^{20}\ \si{cm^{-2}}$. The overall model fit, that in XSPEC notation is {\fontfamily{lmtt}\selectfont TBabs*zpowerlw}, has a $\chi^2/$d.o.f. $= 742.83/711$, providing a good starting point for the spectral analysis.
\begin{figure}
	\centering
	\includegraphics[width=1\linewidth]{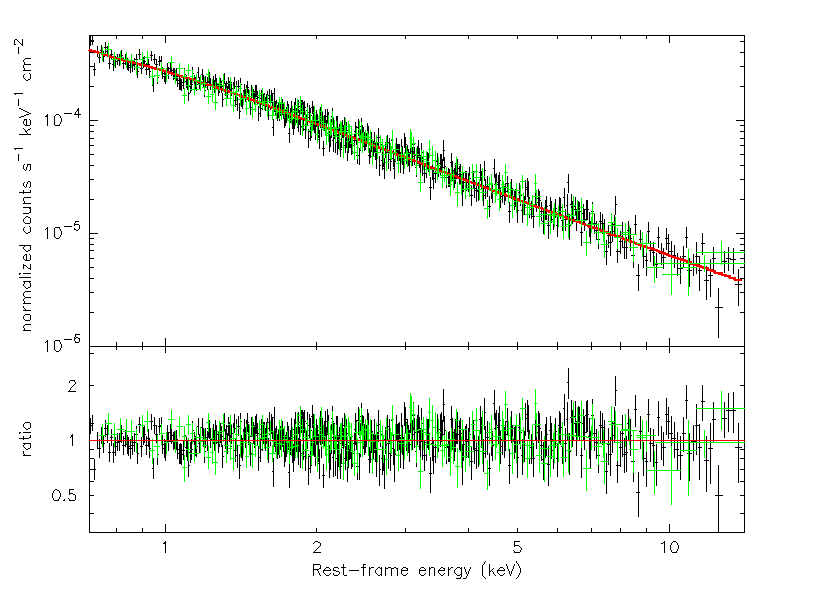}
	\caption{X-ray spectrum (upper panel) and data to model ratio (lower panel) in, using the XSPEC notation, {\fontfamily{lmtt}\selectfont TBabs*(zpowerlw+zpowerlw)} for pn and MOS data (in black and green, respectively). Model is shown with the red line.}
	\label{fig:model1}
\end{figure}
\subsection{Double absorbed power-law}
As discussed in Section \ref{sec:3c215}, this source shows a strong jet emission. In order to explore the possible contribution of the jet in the X-ray spectrum, we added another power-law in the overall model, {\fontfamily{lmtt}\selectfont TBabs*(zpowerlw+zpowerlw)}  in XSPEC notation. The fit provides two different values for the two photon indices, one soft ($\Gamma_1=1.97\pm 0.06$) and one hard $\Gamma_2=1.29 \pm 0.15$), with $ \chi^2/$d.o.f. $=729.36/710$.  From now on, we fixed the jet power-law component to its best-fit value because having two free power-law slopes was causing problems when performing spectral fitting, as the fit routine was often switching between the two slopes. While the first power-law is consistent with the one expected from inverse Compton scattering in the disk-corona system, the second one is similar to the one observable in Flat Spectrum Radio Quasars (FSRQs), that conversely are dominated by the jet. The presence of both corona and jet emission can be expected in the class of broad-line radio galaxies, given their intermediate inclination \citep{grandi2006, kataoka2011, tazaki2013,bostrom2014}. Nonetheless, the jet component is subdominant, with a luminosity $\lesssim 25\%$ of the corona in the  $0.5-10\ \si{keV}$ spectrum, with luminosities in cgs units $\log L_{jet}^{0.5-2 \si{keV}}= 43.81_{-0.10}^{+0.07}$ and  $\log L_{jet}^{2-10 \si{keV}}= 44.35_{-0.10}^{+0.07}$ smaller than the one related to the corona emission, i.e. $ \log L_{corona}^{0.5-2 \si{keV}}= 44.44 \pm 0.02$ and 	$ \log L_{corona}^{2-10 \si{keV}}= 44.53 \pm 0.05$. The significant improvement of the $ \chi^2$ with respect to the single power-law model ($\Delta\chi^2=13.47$ with one d.o.f. less, corresponding to $ P_{jet}=99.97\%$) leads us to include the presence of the jet emission in the X-ray spectrum of 3C 215.\\

\subsection{Modeling the emission and absorption features}
In the ratio plot in Fig. \ref{fig:model1} there are hints of an emission line at $\sim 6.5\ \si{keV}$, likely associated with Fe K emission line, and a possible absorption feature at  $\sim 9\ \si{keV}$. We modeled these emission and absorption features using Gaussian profiles in the source rest frame,  in XSPEC {\fontfamily{lmtt}\selectfont zgauss}, with width $\sigma_{g}$ and peak energy $E_{g}$. For the emission line, the fit of the overall model, that in XSPEC notation is {\fontfamily{lmtt}\selectfont TBabs*(zgauss+zpowerlw+zpowerlw)}, provided the parameter values $\  E_g=6.53^{+0.11}_{-0.08}\ \si{keV}$, $\ \sigma_g<0.41\ \si{keV}$, photon indices $\Gamma_1=1.95\pm0.06$ and  $\Gamma_2=1.29$ and equivalent width $\si{EW}_g=74_{-30}^{+28}\ \si{eV}$, the latter obtained with the XSPEC command {\fontfamily{lmtt}\selectfont eqwidth}. 
Using the Fisher test we derive a confidence level of $P_{em}=98.7 \%$ for the detection of the emission line. This model provides $\chi^2/$d.o.f. $= 720.52/707$. We carried out the same analysis also for the possible absorption line with fixed width, showing $\chi^2/$d.o.f. $= 716.19/705$ while the relative Fisher test provided a probability $P_{abs}=89 \%$, which led us to discard the presence of this absorption feature on statistical grounds. The best-fit values of the parameters are shown in Table \ref{table:model3}. 
\begin{figure}
	\centering
	\includegraphics[width=1\linewidth]{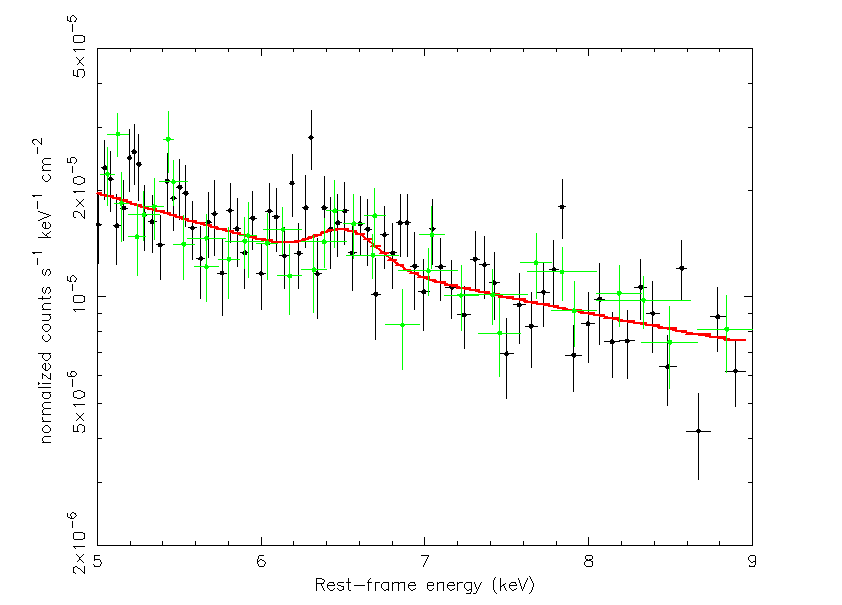}
	\caption{Zoomed pn and MOS data (in black and green, respectively) and model (in red) in the $5-9\ \si{keV}$ band including a Gaussian emission line at $E\simeq 6.53\ \si{keV}$ in model {\fontfamily{lmtt}\selectfont TBabs*(zgauss+zpowerlw+zpowerlw)}, using XSPEC notation.}
	\label{fig:model3lines}
\end{figure}
\begin{table}
	\caption{Best fit-value of the parameters for the model {\fontfamily{lmtt}\selectfont TBabs*(zgauss+zpowerlw+zpowerlw)}.}
	\label{table:model3}
	\begin{center}
		\begin{tabular}{c c c} 
			\hline\hline\\
			Parameter & Value & Units  \\
			\hline\hline
			\\
			Galactic absorption&&\\[1.5ex]
			\hline\\
			$N_H*$&$ 3.50 $& $\times 10^{20}\ \si{cm^{-2}}$\\[1.5ex]
			\hline\hline
			\\
			Power-law 1 (corona)&&\\[1.5ex]
			\hline\\
			$\Gamma$&$1.95\pm0.06$  &- \\[1.5ex]
			$N_{pl}$&$4.11_{-0.17}^{+0.19}$&$\times10^{-4}\ \frac{\si{counts}}{\si{cm^2}\ \si{s}\ \si{keV} }$ \\[1.5ex]
			$z*$& $0.412$ & - \\[1.5ex]
			\hline\hline
			\\
			Power-law 2 (jet)&&\\[1.5ex]
			\hline\\
			$\Gamma*$&$1.29$&- \\[1.5ex]
			$N_{pl}$&$8.15_{-2.06}^{+1.77}$&$\times10^{-5}\ \frac{\si{counts}}{\si{cm^2}\ \si{s}\ \si{keV} }$ \\[1.5ex]
			$z*$& $0.412$ & - \\[1.5ex]
			\hline\hline
			\\
			Gaussian emission line&&\\[1.5ex]
			\hline\\
			$E_{g}$&$6.53_{-0.08}^{+0.11}$ &$\si{keV}$ \\[1.5ex]
			$\sigma_{g}$&$<0.41$&$\si{keV}$
			\\[1.5ex]
			EW$_{g}$&$74_{-30}^{+28} $& $\si{eV}$
			\\[1.5ex]
			$N_{g}$ &$1.87_{-0.73}^{+0.99}$& $\times10^{-6}\ \frac{\si{counts}}{\si{cm^2}\ \si{s}\ \si{keV} }$  \\[1.5ex]
			\hline\hline
		\end{tabular}
	\end{center}
	\captionsetup{{labelformat=empty}}
	\captionsetup{name=Notes}
	\caption{\textbf{Notes.} $N_{pl}$ is the normalization of the power-law, i.e. the flux at $1\ \si{keV}$. $N_{g}$ is the normalization of the Gaussian profile, i.e. the total flux of the line. The starred parameters are the ones kept frozen during the fit. Upper and lower limits are calculated at the 90\% confidence level.}
\end{table} 
\subsection{Reflection from ionized material}
The peak energy $E > 6.4\ \si{keV}$ of the iron emission line  suggests that the reflector is made by different ionized iron species, and this phenomenon can be interpreted as a reflection from ionized gas in the accretion disk. To account for this process we use the additive model {\fontfamily{lmtt}\selectfont xillver} \citep{garcia13},that can describe the reflected spectrum of the radiation which is reprocessed in an ionized accretion disk, requiring that the incident flux is generated by a power-law spectrum, like the Comptonized emission from an hot corona in the AGN environment.\\ In this model it is introduced the ionization parameter $\xi=L_{ion}/nr^2$, where $L_{ion}$ is the luminosity of the incident ionizing radiation in the energy range $E=13.6\ \si{eV}-13.6\ \si{keV}$, while $n$ and $r$ are the density of the reflector and its distance from the ionizing source, respectively. This model is also dependent on the inclination angle of the line of sight $i$. We kept this value fixed at $i=30$\textdegree, which is a typical value for a BLRG, since the fit is unable to constrain it if it is set free to vary.\\
In addition, we used a Gaussian smoothing profile with an energy independent width, in XSPEC {\fontfamily{lmtt}\selectfont gsmooth}, in order to consider the possibility of velocity broadening of the line in addition to the observed broadening due to the presence of emission from a range of iron ions. The overall model fit, in XSPEC  {\fontfamily{lmtt}\selectfont TBabs*(zpowerlw + gsmooth*xillver)}, provides a width upper limit $\sigma_{ion}<82\ \si{eV}$, ionization $\log\xi=3.92_{-0.15}^{+0.13}$ and corona photon index $\Gamma=1.82_{-0.03}^{+0.02}$. The value $\chi^2/$d.o.f. $= 728.75/709$ is comparable with the one of the double power-law model.
\subsection{The relxill model}\label{subsec:relxill}
We find an excess of width of the emission line considering both phenomenological and {\fontfamily{lmtt}\selectfont xillver} models, with a relatively large FWHM of $\sim$10,000 km/s which would suggest a location relatively close to the SMBH.   
Here, we test the hypothesis that the width of the line is due to relativistic effects from reflection onto accretion disk using the model {\fontfamily{lmtt}\selectfont relxill} \citep{garcia&dauser2014a,dauser&garcia2014b}, that takes into account special and general relativity effects in shaping the reflected emission line profile. This model introduces new parameters, namely the dimensionless spin parameter $a=Jc/GM^2$ (kept fixed at the maximum value $a=0.998$), the distance between the inner radius of the reflector and the SMBH $R_{in}$, the distance of the outer radius $R_{out}$ (kept fixed at the standard value $R_{out}=10^3\ R_g$) and the reflection fraction $R_f$, i.e. the ratio between the reflected and direct emission. The best-fit values relative to this model, in XSPEC nomenclature {\fontfamily{lmtt}\selectfont TBabs*(zpowerlw + relxill)}, are $\Gamma_1=1.93\pm 0.02$, $\log\xi=2.31_{-0.27}^{+0.37}$, $R_{in}> 38\ R_g$, $R_f=0.09\pm 0.04$ with $\chi^2/$d.o.f. $= 722.47/708$, showing an improvement of $\Delta\chi^2=6.28$ with 1 d.o.f. less and confidence level $P_{relxill}= 98 \% $ with respect to the {\fontfamily{lmtt}\selectfont xillver} model. The best-fit value of the parameters are shown in Table \ref{table:model5}. We decided to choose {\fontfamily{lmtt}\selectfont TBabs*(zpowerlw + relxill)} as our best-fit model, well characterizing the X-ray spectrum of 3C 215. 
\addtocounter{table}{-1}
\begin{table}
	\caption{Best fit-value of the parameters for the model {\fontfamily{lmtt}\selectfont TBabs*(zpowerlw + relxill).}}
	\label{table:model5}
	\begin{center}
		\begin{tabular}{c c c} 
			\hline\hline\\
			Parameter & Value& Units  \\
			\hline\hline
			\\
			Galactic absorption&&\\[1.5ex]
			\hline
			\\
			$N_H*$&$ 3.50 $& $\times 10^{20}\ \si{cm^{-2}}$\\[1.5ex]
			\hline\hline
			\\
			Relativistic effects&&\\[1.5ex]
			\hline\\
			$\Gamma$&$1.91 \pm 0.02$&- \\[1.5ex]
			$z*$& $0.412$ & - \\[1.5ex]
			$N_{rel}$&$4.77\pm 0.11$&$\times10^{-6}\ \frac{\si{counts}}{\si{cm^2}\ \si{s}\ \si{keV}}$\\[1.5ex]
			$\log\xi$&$2.31_{-0.27}^{+0.37}$& $ \si{erg \ s^{-1} \ cm} $ \\[1.5ex]
			$a*$&$0.998$&-\\[1.5ex]
			$R_{in}$&$>38$& $R_g$ \\[1.5ex]
			$R_{out}*$&  $10^3$ & $R_g$ \\[1.5ex]
			$R_f$&$ 0.09\pm 0.04 $&-\\[1.5ex]
			$i*$&$30$&\textdegree\\[1.5ex]
			$\beta*$ & $3$ & - \\[1.5ex]
			$E_{cut}*$ & $300$ & $\si{keV}$  \\[1.5ex]
			$A_{Fe}*$ & $1$ & $A\ensuremath{_\odot}$  \\[1.5ex] 
			\hline\hline 
			\\
			Power-law 2 (jet)&&\\[1.5ex]
			\hline\\
			$\Gamma*$&$1.29$&- \\[1.5ex]
			$N_{pl}*$&$8.15$&$\times10^{-5}\ \frac{\si{counts}}{\si{cm^2}\ \si{s}\ \si{keV} }$ \\[1.5ex]
			$z*$& $0.412$ & - \\[1.5ex]
			\hline\hline
			\\
		\end{tabular}
	\end{center}
	\captionsetup{{labelformat=empty}}
	\caption{\textbf{Notes.} $N_{rel}$ is the normalization in relxill \citep{dauser2016}. $E_{cut}$ is the energy of the observed cut-off. $A_{Fe}$ is the reflector iron abundance in solar units. $\beta$ is the coronal emissivity slope, which is parameterized as a single power-law. The starred parameters are the ones kept frozen during the fit. Upper and lower limits are calculated at the 90\% confidence level.}
\end{table}
\subsection{Neutral reflection from cold material}
Despite establishing the best-fit model for the spectrum of 3C 215 in Sec. \ref{subsec:relxill} , we tested an additional model component in order to verify the presence or absence of neutral Compton reflection from a distant, cold and optically thick material (such as the putative AGN torus). We used the model {\fontfamily{lmtt}\selectfont pexmon} \citep{nandra2007}. We added this model as a new component to the best-fit model shown in Table \ref{table:model5}. Since we are interested in the neutral Fe K$\alpha$ ($E= 6.4\ \si{keV}$) equivalent width, we freeze all the parameters except for $R_{pex}$ in order to derive limits on the presence of neutral reflection. We obtained a value $R_{pex}\simeq 0$ for the neutral reflection fraction. We also computed the Fe K$\alpha$ EW modeling the line with a Gaussian profile, fixing the energy peak $E=6.4\ \si{keV}$ and line width $\sigma=0$ and adding this new component to the ionized iron emission line model. We obtained an upper limit EW$_{\si{K\alpha}} <26\ \si{eV}$. In order to explore the jet effect on this estimate, we performed the same calculation eliminating data and the second power-law in the model (the one describing the jet emission), and we obtained that EW$_{\si{K\alpha}} <41\ \si{eV}$ . The upper limit is computed by fixing the line intensity with its best-fit value plus the associated intensity error. This is an indication for absent or weak neutral reflection features, whose implications for the putative AGN torus and SMBH feeding will be discussed in Sec. \ref{discussion}.

\section{Results and discussion}\label{discussion}
The best-fit model we found in the spectral analysis in Sec. \ref{sec:spectralanalysis} describes the X-ray spectrum of 3C 215 as an absorbed double power-law with a tentative detection of an iron emission line peaking at $E\sim 6.53\ \si{keV}$. The shape and ionization of the emission line is affected by the interaction with an ionized reflector and relativistic blurring, that are both taken into account in the model.\\
We computed the ionizing continuum luminosity $L_{ion}\simeq 1.38 \times 10^{45}\ \si{erg\ s^{-1}}$, modeling the ionizing spectrum as a single power-law between $E=13.6\ \si{eV}-13.6\ \si{keV}$ without the contribution of the jet and using the XSPEC command {\fontfamily{lmtt}\selectfont clumin}.
We also calculated the X-ray luminosity $L_X=(5.49 \pm 0.13) \times 10^{44}\ \si{erg\ s^{-1}}$, defined as the spectrum luminosity in an energy range between $E=2-10\ \si{keV}$ considering both the contributions of the corona and the jet.\\
Previous observation of this source in the optical/UV band provided an estimate of the central SMBH mass $\log(M_{BH}/ \si{\solarmass}) =8.92\pm 0.5$ \citep{tang2012}. From this value we are able to estimate the Eddington luminosity $L_{Edd}=1.3 \times 10^{38} M_{BH}/M_{\odot}\ \si{erg/s}$ and the gravitational radius $R_g= GM_{BH}/c^2$, which will be used as an unit of measure for distances in the nuclear region on this AGN. Their values in log scale and cgs units are $\log L_{Edd}=47.03\pm0.5$ and $\log R_g=14.9\pm0.5$, respectively.\\
Merging the information obtained from our analysis with the ones present in literature it is possible to derive a more detailed description of this AGN and of its nuclear environment.

\subsection{Jet contribution in the X-ray spectrum}
The $0.5-10\ \si{keV}$ spectrum of 3C 215 shows a jet contribution in the form of a flat power law ($\Gamma \sim 1.29$) together with a second steep power law ($\Gamma \sim 1.95$) likely associated with the Inverse Compton emission originated in the corona. As discussed in the previous section, the contribution of the jet power law to the overall X-ray spectrum is subdominant with respect to the corona emission ($L_{jet} \lesssim 25 \% $), in accordance with other BLRG observations \citep{grandi2007}. This result is also in agreement with 3C 215 mixing parameter estimate $\eta \simeq 0.034$ \citep{kataoka2011}, computed through SED fitting and defined as the ratio between non-thermal and thermal emission.\\
The jet component seems to connect this source to blazars, and in particular to FSRQs, which among the blazar class have on average harder photon indices in the X-ray band \citep{sambruna1994, comastri1997, sambruna1997,fan2012}. Moreover, the observed optical broad emission lines \citep{tang2012}, the high luminosity and flat radio spectrum (shown in 3C 215 SED reported in the NASA/IPAC Extragalactic Database) further corroborates this hypothesis.\\
The joint observation of a steep Seyfert-like spectrum and a FSRQ spectrum in the X-ray band leads us to assume an intermediate inclination angle for this source, as expected for BLRGs.\\ Nonetheless, these particular sources strongly challenge the Unified Scheme, since there are observations of BLRGs with a visible jet contribution (e.g. \cite{bostrom2014}) while other BLRGs do not exhibit the jet component in the X-ray band (e.g. \cite{ronchini2019}), leading to assume that also in the BLRG class there exist other sub-classes with different observative features that may not be explained only by different observational angles, since it is still unclear at which inclination angle a BLRG stops to be "Blazar-like" and starts to be more "Seyfert-like".
\subsection{Ionized reflector}\label{sec:distant reflector}
Our data analysis of 3C 215 shows that the reflector, which interacts with the ionizing primary radiation from the hot corona, is placed at distance $R_{in}\gtrsim 38\ R_g$ from the SMBH, that is relatively far compared to the ISCO value of $R_{in}\sim 1\ R_g$ expected for an AGN with a SMBH spin $a\sim1$. It is also ionized, with an ionization parameter $\xi \simeq L_{ion}/nR^{2}_{in} \simeq 204 \ \si{erg \ s^{-1} \ cm}$. Inverting the relation, we can obtain an upper limit for the reflector particle density $n\lesssim  10^{10}\ \si{cm^{-3}}$.\\ The density estimated in this calculation is low for an internal region of the accretion disk, whose density is expected to be of the order of $\sim 10^{15}\ \si{cm^{-3}}$. The density of the reflector seems to be more in agreement with the BLR, with a typical density of $\sim 10^{10}\ \si{cm^{-3}}$, or an outer ionized layer of the accretion disk.\\
In previous observations in the optical/UV band of 3C 215 there were hints of  a stratified BLR with a disk-like geometry for the region producing the C$_{\si{IV}}$ emission line \citep{labita2006}. \cite{tang2012} provided the FWHM of some broad optical emission lines of this source, like C$_{\si{IV}}$ and H$_{\beta}$, with values $\si{FWHM(C_{IV})}= 5605_{-85}^{+90}\ \si{km/s}$ and $\si{FWHM(H_{\beta})}= 6760_{-505}^{+500}\ \si{km/s}$, respectively.\\
We observe that the FWHM(H$_{\beta}$)  is slightly larger than the C$_{\si{IV}}$ one. From standard BLR stratification models, we would expect the opposite, with the more ionized line produced more internally hence with an higher FWHM, related to the velocity dispersion. This evidence can be explained considering that, since the line of sight is tilted with respect to the equatorial plane, we probably observe an H$_{\beta}$ outflowing source, that has an higher velocity dispersion along the line of sight hence an higher FWHM, even if the source projected on the equatorial plane is more external than the C$_{\si{IV}}$ one, which in contrast has a disk-like geometry.\\ 
Considering the reverberation mapping relation $R_{BLR}=GM_{BH}/\si{FWHM}^2$, the disk-like portion of the BLR can cover the accretion disk surface at distances of $\gtrsim 10^2\ R_g$ thus contributing to the reflection phenomenon that generates the reflected spectrum of 3C 215 in the X-ray band.\\
For distances $R\lesssim 38\ R_g$ the ionizing flux from the hot corona can fully ionize elements lighter than iron, thereby excluding  emission lines in the optical/UV band. This may, at least partially, explain why there is no presence of observable reflection features in the SMBH neighborhood.\\
At the distance of $R_{BLR}$, where the bulk reflection from the BLR is taking place, we expect a reflector less ionized with respect to the reflection in the inner portion of the disk, where $\log\xi = 2.31_{-0.27}^{+0.37}\ \si{erg \ s^{-1} \ cm}$. From the calculations, considering $n \sim 10^{10}\ \si{cm^{-3}}$ and $R_{BLR}\sim 4 \cdot 10^2\ \si{R_g}$, calculated as the mean value of the two emission lines, we obtain:
\begin{equation}
\xi_{BLR}=\frac{L_{ion}}{nR^{2}_{BLR}} \sim 1.6\ \si{erg \ s^{-1} \ cm}\ll 204\ \si{erg \ s^{-1} \ cm} 
\end{equation}
These evidences suggest a standard optically thick and geometrically thin structure in the outer part of the accretion disk of 3C 215, while the inner part could be a truncated disk or a composite disk, where the inner part is radiatively inefficient, i.e. in an ADAF like state, or alternatively a puffed-up slim disk.

\subsection{Long-term time variability of the X-ray spectrum}\label{sec:xrayvariab}
In order to better describe this AGN, it is important to compute the Eddington parameter $\lambda=L_{bol}/L_{Edd}$, where $ L_{bol} $ is the bolometric luminosity. We were able to estimate $ L_{bol} $ starting from $ L_{X} $ through the bolometric correction $K_X=L_{bol}/L_{X}$ parametric equation derived by \cite{duras2020}. Using this procedure we computed the Eddington parameter $\lambda$ for our observation and for the previous X-ray observations of 3C 215 in the energy band $E=0.5-10\ \si{keV}$ carried out by \cite{reeves2000} (R\&T00) and \cite{hardcastle2006}(H06) which, using observations of different satellites, managed to estimate the X-ray luminosity $L_X$. The results are summarized in Table \ref{tab:xrayvar}.\\
\addtocounter{table}{-1}
\begin{table}
	\caption{Information and parameter values concerning the observations of 3C 215 in \cite{reeves2000} (R\&T00), in  \cite{hardcastle2006} (H06) and in our spectral analysis.}
	\label{tab:xrayvar}
	\begin{center}
		\begin{tabular}{c c c c} 
			\hline
			\hline
			\\
			&R\&T00 & H06 & This work \\
			\hline
			\\
			Telescope &  ASCA & Chandra & XMM \\[2ex]
			Obs. year &  $ 1995 $  &  $2004$ & $2012$\\[2ex]
			$L_X[10^{44}\si{\frac{erg}{s}}]$ & $17.20$ & $6.92$ &$5.49 \pm 0.13$\\[2ex]
			$K_X$ & $34.63$ & $26.36$ &$24.97 \pm 0.22$\\[2ex]
			$L_{bol}[10^{46}\si{\frac{erg}{s}}]$ & $5.96$ & $1.82$ &$1.37 \pm 0.04$ \\[2ex]
			$\lambda$ & $0.54$ & $0.17$ &$0.13^{+0.28}_{-0.09}$\\[2ex]
			\hline\hline
		\end{tabular}
	\end{center}
\end{table}
Comparing these multi-epoch estimates, we observed variability of the X-ray spectrum on time-scales of $\sim8$ years. $L_X$ decreases, producing a recent spectrum dimmer with respect to the one observed $\sim 17$ years before our observation. Evidences of this kind of variability in the X-ray spectrum of AGNs were already found in Seyfert 1 galaxies \citep{noda2014,noda2016}, with time scales from weeks to years.\\
There is no evidence of a jet contribution in the long-term variability of this source, both because it is not dominant in the X-ray band and because past observations where not precise enough to evidence this feature.\\
This phenomenon can be attributed to a change of the accretion rate \citep{noda2014}. An efficiently accreting SMBH is very luminous and forms an optically thick and geometrically thin disk up to the ISCO and a subdominant, if present, internal ADAF state. If the accretion rate and consequently the luminosity decrease, as in this case, the ADAF state is no more subdominant in the internal region, the disk emissivity drops and the dominance of the hot corona in the overall X-ray emission produces an harder spectrum.\\ 
With the same principle \cite{noda2018} explain the changing-look phenomenon of the source \object{Mrk 1018}, which in a time range of $\sim8\ \si{yrs}$ switches from a Seyfert 1 galaxy to a Seyfert 2. They state that the variation in the accretion rate is driven mainly by two combined factors, previously observed in BH binary systems:\\
\begin{itemize}
	\item[i.] Evaporation/condensation of the inner disk in an advection-dominated hot flow, which causes a decrease/increase of the luminosity of a factor of $\sim2-4$;
	\item[ii.] Thermal front propagation due to the H-ionization instability in the disk.
\end{itemize}
In the case of 3C 215, we observe a variability of almost a factor of $\sim 3$ from 1995 to 2012, which is more comparable to the classic X-ray variability of AGNs rather than a changing-look phenomenon, where luminosity drops by at least an order of magnitude.\\ 
Therefore, the observational evidences of 3C 215 suggest an evolution of the inner structure of the AGN, consistent with the luminosity drop caused by the evaporation of the inner thin disk, as proposed by \cite{noda2018}.\\
However, we must note that the Eddington parameter estimated through our analysis is not completely in agreement with the values expected by this scenario, which is expected to be $\lambda\sim 1\%$. This leads us to believe that, with respect to the other sources, in 3C 215 the contribution of the standard outer disk is more relevant, questioning the presence and the importance of the ADAF in the inner disk.
\subsection{Lack of neutral absorption and reflection from the torus}\label{sec:torus}

We want to underline the role of the putative obscuring torus that surrounds the nuclear regions of 3C 215. It is typically made by optically thick material that interacts with the X-ray emission from the nuclear region through absorption and Compton scattering, giving important information about its geometry. Because of its distance from the SMBH, typically of $\sim$pc scale, it is too cold to be ionized, hence we test its presence analyzing the neutral reflection component of the spectrum. From the data analysis we find an upper limit for the equivalent width of the neutral Fe K$\alpha$ line without jet contribution EW$_{\si{K\alpha}}\lesssim41\ \si{eV}$, remarking a weak interaction between the primary X-ray radiation and the torus, that gives a minimal contribution to the overall X-ray spectrum. Theoretical models associate to such a weak or absent iron line feature a disrupted and inefficient obscurer with a patchy geometry, low column density $N_H$ (with an upper limit $N_H < 5\cdot 10^{22}\ \si{cm^{-2}}$) \citep{ikeda2009,murphy2009} and poor presence of dust \citep{gohil2015}. This hypothesis is also supported by the value of the reflection parameter $R_{pex}\simeq0$ obtained in the spectral analysis, describing a torus which covers only a small fraction of the space surrounding the AGN. Obscuring material can represent the main matter reservoir for accreting SMBHs, triggering feedback through nuclear outflows or jet emission, as in the case of RL AGNs. For high Eddington parameters $\log\lambda> -1.5$ (as in 3C 215) the covering fraction of the torus is small ($\sim40 \%$) because a large fraction of that material was and maybe it is still being accreted by the central SMBH, bringing to high luminosity values \citep{ricci2017}.
Seyfert 1 galaxies show a similar obscuration but absence of jet emission, possibly substituted by nuclear outflows, producing Fe K$\alpha$ EW of the order of $\sim 53\ \si{eV}$ \citep{shu2010}.\\
The characteristics described in this scenario may appear in disagreement with what was observed in the IR band of 3C 215 from \cite{runnoe2013}. They found a covering fraction related to the MIR radiation $c=64 \%$, which is very large considering that the torus does not show significant presence of dust, which permits the interaction between the torus and the IR flux. An interpretation of this discrepancy could be the presence of a second obscuring component such as a polar dusty outflow, luminous in the IR band \citep{honig2012,ramosalmeida2017}. This wind can account for the large covering factor estimated by previous studies. Nonetheless, it does not contradict our results regarding the obscuring torus obtained with our X-ray analysis.
\subsection{Slim disk?}
As already discussed in Sec. \ref{sec:distant reflector}, the lack of reflection features in the internal region of this radio-loud quasar can possibly be in agreement with a double-state of the accretion disk, where the outer portion shows a standard Shakura-Sunyaev geometry \citep{shak73}, while the internal one ($R_{in}\lesssim 38\ \si{R_g}$) is radiatively inefficient and geometrically thick, in order to favor the production of a jet. Nonetheless, ADAF states are present only in the disk of low accreting SMBHs. If the Eddington parameter is high ($\lambda\gtrsim 0.3$) the disk becomes thick enough to make advection an efficient cooling mechanism, hence leading to a drop of the disk emissivity. This disk state, related to highly accreting SMBHs, is called slim disk \citep{abramowicz1988,abramowicz&fragile2013,yuan2014} and within the uncertainties it is in agreement with the Eddington parameter $\lambda=0.13^{+0.28}_{-0.09}$ here estimated.\\ According to this scenario, the presence of a composite accretion disk leads to an anisotropic radiation field, that interacts differently with the various components. Consequently, a self-shadowing phenomenon can produce two dynamically distinct zones of the BLR, with one of them possibly composed by H$_{\beta}$ outflowing clouds  \citep{wang2014}, in agreement with the previous BLR discussion of 3C 215. The higher the accretion rate, the greater is the self-shadowing effect, hence we do not expect it to be very intense in this source, since it is usually related to super-Eddington SMBHs.\\
In relation with this local disk emissivity drop, one could expect also a drop of the interaction between the UV incident flux and the torus, leading to a decrease of the reprocessed IR flux and to a torus characterization bias, since it would appear as an inefficient obscurer. 
Nonetheless, these features are also observed in the X-ray band of 3C 215, hence they are not due to only the self-shadowing bias. In addition, to account for the intense IR emission of this source we proposed the presence of a second dusty component. In this scenario, this reprocessing component could be constituted by an outflowing BLR, possibly dust rich \citep{czerny2011,baskin2018}.\\
Although this scenario is able to describe several observational features of 3C 215, it does not provide an explanation of the particular geometry of the jet emission observed in this source, remarking the necessity of a possible new scenario, discussed in the next section. 
\subsection{A SMBH binary system?}
A third speculative scenario to explain our X-ray and previous multi-wavelength observations of 3C 215 is one involving the presence of a second SMBH, forming a SMBH binary system (SMBHB) \citep{begelman1980,derosa2019}.\\
Different motivations led us to consider the possibility of a SMBHB in describing this source, like hints of a previous major merger event involving 3C 215 host galaxy (as we discussed in Sec. \ref{sec:3c215}), the twisted and knotted radio jet emission in this AGN, possibly due to the formation of warped accretion disks \citep{bardeen&petterson75,papaloizou1983,nayakshin2005,ulubay-Siddiki2009} and/or direct interaction of the binary with the jet producing region \citep{lawrence2010,tremaine2014}, the perturbed shape of the obscuring torus and the relatively high accretion rate of this source. All these features can be explained by the presence of a putative secondary SMBH, that perturbs the space surrounding its orbit. If the distance between the two SMBHs would be at sub-pc scales, this can possibly also explain the lack of material, hence reflection features, in the internal portion of the accretion disk at distances $\lesssim38\ \si{R_g}$, as observed in the X-ray analysis.\\
Nonetheless, SMBHBs are very elusive and hard to recognize since they can not be spatially resolved, such that a large fraction of binary system detection was serendipitous. The state-of-the-art methods to select these systems are multi-epoch spectroscopic observations, multi-band follow-up and cross-correlation analyses, that are able to show some signatures of the periodic motion of the binary. Therefore, the current data available of 3C 215 are not enough to
firmly select it as a strong SMBHB candidate. Nonetheless, there exist two main SMBHBs selection methods that rely only on single-epoch spectroscopic observations of broad emission lines (BELs) in the optical/UV band.\\ The first method consists in detecting periodical and intense Doppler shift $\Delta v$ of BEL peaks produced in the BLR, linked to the internal orbital motion in the same way spectroscopic binary stars are observed \citep{bogdanovic2008, eracleous2012,decarli2013}. $\Delta v$ can be computed by the difference between Mg$_{\si{II}}$ and [OIII] peak velocities, the latter having with good approximation the same velocity of the host galaxy \citep{ju2013,wang2017}. For 3C 215 the estimate $\Delta v \simeq 915\ \si{km\ s^{-1}}$ \citep{tang2012} is very high compared to usual values (e.g. \cite{wang2017}) and close to the selection threshold $\lvert\Delta v\lvert \ \gtrsim 1000\ \si{km\ s^{-1}}$.\\
The second method is based on the fact that during binary orbital decay, the BLR starts to emerge from the Roche lobes of the binary, producing outflows of material and consequently the BLR erosion. Since this process affects more the external region of the BLR, one could expect that low ionization emission lines should suffer a significant flux decrease with respect to the high ionization ones, which are produced more internally. Interestingly, the BLR erosion is consistent with the hypothesis of an outflowing external portion of the BLR proposed in Sec. \ref{sec:distant reflector}. \cite{montuori2011,montuori2012} found that the ratio between the fluxes of the  C$_{\si{IV}}$ line ($F_{CIV}$) and of the Mg$_{\si{II}}$ line ($F_{MgII}$) increases with increasing orbital period $P$, hence this ratio constitutes a good binary diagnostic.\\
To be selected as SMBH binary candidates  sources have to satisfy the relation $F_{MgII}/F_{CIV}\lesssim 0.1$, with a threshold smaller with respect to the common value which is $F_{MgII}/F_{CIV} \sim 0.3-0.4$.\\
3C 215 shows a flux ratio $F_{MgII}/F_{CIV}=\  0.14 \pm 0.04$,  consistent with the selection requirement. However, these two analysis can only select 3C 215 as a potential weak SMBHB candidate, being them not able to provide a clear and unequivocal evidence of the presence of the binary through the detection of orbital modulations.

\section{Conclusions}

We performed the most accurate X-ray spectral analysis of the broad line radio galaxy 3C 215 using high quality data from the XMM-Newton satellite. Comparing the results we obtained with multi-wavelength observations published in the literature we were able to characterize the innermost components of this AGN, modeling their geometry and mutual interactions.\\
This source shows a jet contribution in the X-ray band that can be described as a flat power law, very similarly to FSRQs. The primary X-ray continuum from the corona interacts with a reflector placed  at relatively large distances ($R\gtrsim 38\ R_g$) from the SMBH. Evidences suggest that the reflection component is the accretion disk, which is ionized ($\log\xi \simeq 2.31\ \si{erg\ s^{-1}\ cm}$), and also part of the BLR, that shows a stratified structure, with the C$_{\si{IV}}$ emission line emitter exhibiting a disk-like geometry over the accretion flow and a putative outflowing H$_{\si{\beta}}$ source region. The ionized disk gas infalls towards the SMBH, that accretes its mass with an high sub-Eddington rate ($\lambda=0.13^{+0.28}_{-0.09}$).\\
Spectral features expected from the putative obscuring torus are very weak, if not absent, and this suggests a patchy geometry with a low covering fraction ($\sim40\%$) and reflection parameter ($R_{pex}\simeq 0$). The low \iron equivalent width (EW$<41\ \si{eV}$) suggests a minimal fraction of dust in the X-ray torus structure, in contrast with previous observations in the literature, reporting a covering fraction $c=64\%$ in the MIR band directly linked with a dusty obscurer. This can be explained by the presence of an internal polar dusty wind that participate together with the torus in the nuclear obscuration process.\\
We compared the results of multiple spectral analysis carried out in different years, we have observed an intense variability of the X-ray spectrum of 3C 215, that is becoming fainter (almost a factor of 4) with an overall time step of $\sim17$ years. This variability can be associated with a drop of the Eddington parameter that brings to the evaporation of the inner edge of the accretion disk in an radiatively inefficient geometrically thick accretion flow.\\
We propose three scenarios that could explain these observational properties:
\begin{enumerate}
\item ADAF state in the internal region, which is in accordance with the reflector distance but not with the accretion rate, too high for this structure;
\item Slim inner disk, which has a similar geometrical structure of the ADAF one, it is consistent with 3C 215 accretion regime and accounts for almost all the observational features, except for the unusual jet emission shape;
\item SMBH binary system, which may describe the overall structure of 3C 215 in a good way but lacks on a firm SMBHB detection, making this scenario speculative. 
\end{enumerate}
Some observational features manifested by 3C 215 are common in RL sources, like weak reflection features (e.g.  \cite{bostrom2014,ursini2018b,ronchini2019,chalise2020}).
In contrast, RQ AGNs show an opposite behavior, with a standard accretion disk that extends down to the ISCO, being able to produce strong reflection features like the \iron line and provide good spin parameter estimates \citep{fabian2014,reynolds2020}.\\
Nonetheless, 3C 215 shows an obscuring torus not so different from the ones in RQ AGNs \citep{shu2010,tazaki2013,panessa2016} and its accretion rate $\lambda$ is high compared to other RL AGNs, and more similar to the one expected in RQ sources. In fact, as we discussed in Sec. \ref{sec:torus} these two characteristics are strongly linked. However, the accretion disk geometry, which depends on the accretion rate, must be thick enough to allow matter collimation and jet production \citep{ballantyne2007}. As we discussed, this peculiar geometry can be obtained both with high and low accretion rates, thus suggesting that the important phenomenon that helps jet production can possibly be the disk height \citep{abramowicz&fragile2013} or the mutual rotation direction between the SMBH and the accreting material \citep{garofalo2019}, rather than the accretion rate itself.\\
This observation contributes to the BLRG catalog and, through the hints of jet emission in the X-ray band, allows us to better understand how much heterogeneous these sources are, at least in the X-ray band. Discovering how and why BLRGs can show different observative outputs can be crucial in solving the RL-RQ dichotomy.\\
Unfortunately, with the data obtained through our observation and the previous ones in the literature we can achieve only a weak SMBHB candidate selection, although the selection tests we made for 3C 215 are promising. New multi-epoch spectroscopic observations, multi-band follow-ups and cross-correlation studies are required in order to firmly select 3C 215 as a strong SMBH candidate and to possibly add this source in the new developing SMBHB catalogs, which can be compared with the one related to classical single AGNs. In particular with new observations in the radio band, for example with the SKA facility, it will be possible to resolve the jet source of 3C 215 with high resolution, thereby acquiring new information about the process that bends its jet. Also further time monitoring in the optical/UV band with the Hubble Space Telescope (HST) could provide hints of some kind of time variability in the emission lines, possibly associated with orbital periods of a putative binary system. Last but not least, an X-ray follow-up can further monitor the X-ray variability observed in this source, providing an additional validation of the SMBHB scenario. 

\begin{acknowledgements}
AM and FT thank the anonymous referee for the constructive comments.
This work was performed by AM for his Master Thesis in Physics -
Curriculum Astrophysics at the Tor Vergata University of Rome based on
observations obtained with XMM-Newton, an ESA science mission with
instruments and contributions directly funded by ESA Member States and
NASA. AM and FT thank F. Vagnetti, F. Panessa and G. Bruni for the useful discussions.
\end{acknowledgements}

\bibliographystyle{aa}
\bibliography{Mei}

\end{document}